\documentclass[aps,preprint,onecolumn]{revtex4}%
\usepackage{amsfonts}
\usepackage{amsmath}
\usepackage{amssymb}
\usepackage{graphicx}%
\providecommand{\U}[1]{\protect\rule{.1in}{.1in}}
\newtheorem{theorem}{Theorem}
\newtheorem{acknowledgement}[theorem]{Acknowledgement}

\begin{document}
\title{Neutron interferometry and tests of short-range modifications of gravity}
\author{J. M. Rocha}
\author{F. Dahia}
\affiliation{Department of Physics, Federal University of Para\'{\i}ba - Jo\~{a}o Pessoa
-PB - Brazil }
\begin{abstract}

We consider tests of short-distance modifications of gravity based on neutron interferometry in the scenario of large extra dimensions. Avoiding the non-computability problem in the calculation of the internal gravitational potential of extended sources, typical of models with zero-width brane, we determine the neutron optical potential associated with the higher-dimension gravitational interaction between the incident neutron and a material medium in the context of thick brane theories. Proceeding this way, we identify the physical quantity of the extra dimension model that the neutron interferometry is capable of constraining.

We also consider interferometric experiments in which the phase shifter is an
electric field, as in the test of the Aharanov-Casher effect. We argue that
this experiment, with this non-baryonic source, can be viewed as a test of the
short-range behavior of Post-Newtonian parameters that measure the capacity of
the pressure and the internal energy for producing gravity.

\end{abstract}
\maketitle

\section{Introduction}

Braneworld scenarios \cite{ADD1,ADD2,RS1,RS2}, according to which our ordinary
four-dimensional spacetime is embedded in a higher-dimensional ambient space,
have motivated many recent studies on the modifications of gravity in
short-distance scale \cite{review}.

Compared to the Kaluza-Klein pioneering work on extra dimension formally based
on General Relativity, these new higher-dimensional theories, among which the
ADD model \cite{ADD1} is a prototype, are distinguished by a peculiar feature.
They are based on the fundamental assumption that matter and fields are
confined in the 3-brane (the ordinary three-dimensional space) while gravity
has access to all directions \cite{ADD1,ADD2,RS1,RS2}. The dilution of the
gravitational field through the whole ambient space would be the reason for
the feebleness of gravity in comparison to the strength of the localized
fields and, therefore, could be the explanation for the hierarchy problem,
whose resolution was originally the main motivation for some braneworld models
\cite{ADD1,ADD2}.

In these scenarios, the effects of extra dimensions on the gravitational field
may become significant in a length scale $R$ much greater than the scale in
which the confined fields would feel them directly. Thus, extra dimensions
with a size much larger than the Planck length are phenomenologically feasible
since gravity, differently from what happens to the other fields, is being
empirically tested in submillimeter domains only recently \cite{review,Hoyle}.

One important prediction of the existence of large extra dimensions according
to these braneworld models is a strong amplification of the gravitational
field in short distances $\left(  r<<R\right)  $, implying that the theory
could be experimentally checked. This possibility has motivated the search for
empirical signs of the supposed hidden dimensions in laboratory tests from
areas such as spectroscopy
\cite{specforxdim1,atomicspec,Li,Wang,specforxdim2,molecule,dahia,fractal,newphys,pHebound,5force}%
, neutron interferometry \cite{frank,pokoti,greene,neutrons,snow}, torsion
balance experiments \cite{Hoyle} and those involving Casimir effect
\cite{lamoreaux,sushkov,bezerra}.

Regarding experimental tests, modifications of gravity are usually
parametrized by means of an additional Yukawa-like potential energy of the
form $\left(  \alpha GMm/r\right)  \exp(-r/\lambda)$, where $M$ and $m$ are
the masses of interacting particles and $G$ is the Newtonian gravitational
constant. Experiments from diverse areas put upper bounds on the amplification
factor $\alpha$ in different ranges of the length scale parameter $\lambda$
\cite{review}.

The Yukawa parametrization is very useful because it may encompass
modifications of gravity with different theoretical origins \cite{extensions,
models, stelle}. The ADD model predicts a correction of the same type for the
external gravitational potential produced by a particle in the domain of large
distances $r>>R$ \cite{Kehagias}. In this case, the parameter $\alpha$ is
proportional to the number of hidden dimensions $\delta$. The exact relation
depends on the topology of the supplementary space and on the length scale at
which the stabilization of its volume takes place \cite{Kehagias,adelbergREV}.
In its turn, in short distances, the gravitational potential is expected to
exhibit a power-law behavior, i.e., it should be proportional to
$(R/r)^{\delta+1}$ \cite{ADD1}.

It is important to remark that both Yukawa and the power-law approximations
are valid for pointlike particles. In configurations where the particles' wave
functions overlap, the internal gravitational potential of the source should
be considered. It happens that inside an extended source, the gravitational
potential is not computable in a scenario of zero-width brane with $\delta
\geq2$ \cite{effbrane,colliders,dahia}. One way to circumvent this difficulty
is to consider a thick brane model \cite{rubakov,kaplan,fermions,split},
which, allows us, for instance, to estimate the influence of hidden dimensions
in the energy shift of $S-states$ in the Hydrogen-like atoms
\cite{dahia,dahia2}.

Experiments based on the interferometry of neutrons constitute important tests
of non-standard gravity and provide some of the most stringent upper bounds on
the Yukawa parameter $\alpha$ in the range between $10^{-12}%
%TCIMACRO{\unit{m}}%
%BeginExpansion
\operatorname{m}%
%EndExpansion
$ and $10^{-9}$ $%
%TCIMACRO{\unit{m}}%
%BeginExpansion
\operatorname{m}%
%EndExpansion
$ \cite{neutrons}.

Schematically, in this kind of experiment, incident neutrons are divided into
two beams that follow spatially separated paths and are then recombined to
form an interference pattern in the detected intensity of the neutron flux
\cite{rauch}. A quantum mechanical phase difference is acquired when the
partial beams are subjected to different physical interactions along the two
paths as they have to pass through material plates or to cross regions with
electromagnetic fields, according to the aim of the experiment.

In general, most analyses are based on the Yukawa parametrization. However,
when the neutron is in contact with the medium, its phase is affected by the
neutron optical potential which depends on the internal gravitational
potential of the material. Thus, in order to estimate the effects of hidden
dimensions on the neutron interferometry, we calculate the internal potential
of a solid phase shifter in the context of a thick brane scenario, thus
avoiding the divergence problems related to models with zero-width brane. More
specifically, in section II, we determine the forward scattering length of the
anomalous gravitational interaction between the incident neutron and the
nucleus of the material in the leading order, identifying, thus, the physical
quantity of the extra dimension model that the neutron interferometry is able
to constraint. As expected, this quantity depends on a parameter related to
the localization of the matter in the brane, but, as we shall see, it also
depends on the nuclear model of the atomic nuclei that constitute the material.

One way to surpass this limiting aspect is to consider experiments in which
the phase shifter is a non-baryonic source. An example is the experiment
conceived to test the Aharanov-Casher (AC) effect \cite{AC}, which can be
described as a version of the Aharonov-Bohm effect for an electric neutral
particle \cite{rauch,AC}. In the AC experiment \cite{cimmino}, the phase
difference arises from the interaction between the neutron's magnetic moment
and the electric field as the beams cross the interior of electrostatic chambers.

According to the General Relativity theory, all kinds of energy are capable of
curving spacetime. So, in that experiment, incident neutrons also interact
with the gravitational field produced by the electric field. In section III,
we calculate the additional phase shift due to this interaction and discuss
the possibility of extracting independent constraints for short-distance
modifications of gravity from the AC experiment.

These interferometric bounds can be considered in a more general context of
metric theories and their Post-Newtonian parameters \cite{ppn}. As we shall
see, the constraints put from this non-baryonic source can be seen as limits
for short-distance deviations of two Post-Newtonian parameters that measure
the capacity of the internal energy and pressure to bend spacetime in
comparison to the rest mass of matter.

For a restricted class of metric theories, the interferometric constraints
obtained here can be compared to the bounds extracted from the MTV-G
experiment \cite{mtvg,mtvg1,dahia4} and from the spectroscopy \cite{dahia4}
concerning short-distance modifications of the post-Newtonian $\gamma
$-parameter (related to the curvature of spatial sections of the spacetime).
As we shall see, neutron interferometry establishes the most stringent bounds
for this parameter in the length scale between $1.4\times10^{-7}%
%TCIMACRO{\unit{m}}%
%BeginExpansion
\operatorname{m}%
%EndExpansion
$ and $10^{-4}%
%TCIMACRO{\unit{m}}%
%BeginExpansion
\operatorname{m}%
%EndExpansion
$.

\section{Internal potential of an extended source in thick brane scenario}

According to the ADD-model, the spacetime has a certain number ($\delta$) of
compact spacelike extra dimensions. The background spacetime is flat and
contains a supplementary space with a finite volume, $(2\pi R)^{\delta}$.
Matter and all the standard model fields are confined in the brane. Hence, the
energy-momentum distribution of the localized fields may be described by a
tensor of this kind \cite{colliders}:%
\begin{equation}
T_{AB}=\eta_{A}^{\mu}\eta_{B}^{\nu}T_{\mu\nu}\left(  x\right)  f\left(
z\right)  . \label{Tab}%
\end{equation}
Here we are adopting the following notations: Greek indices run from $0$ to
$3$, and capital Latin indices go from $0$ to $3+\delta$. The ordinary
spacetime coordinates are represented by $x$, while $z$ indicates coordinates
of the compact space. The tensor $\eta_{AB}$ is the Minkowski metric. For an
idealized zero-width brane, $f\left(  z\right)  $ is a Dirac delta
distribution, but, in the case of a thick brane, $f\left(  z\right)  $ is some
regularization of that singular distribution.

The confined fields are the source of a gravitational field in the bulk that
obeys a higher-dimensional version of Einstein equations. In the weak-field
regime, the metric is approximately given by $g_{AB}=\eta_{AB}+h_{AB}$, where
the tensor $h_{AB}$, which describes small perturbations in the geometry,
satisfy the linearized Einstein equations:
\begin{equation}
\square h_{AB}=-\frac{16\pi G_{D}}{c^{4}}\bar{T}_{AB}. \label{linear eq}%
\end{equation}
Here the symbol $\square$ is the D'Alembertian operator associated with the
Minkowski metric with a signature $(-,+,...,+)$ and $\bar{T}_{AB}=\left[
T_{AB}-(\delta+2)^{-1}\eta_{AB}T_{C}^{C}\right]  $. It is important to remark
that the above equation is valid in the harmonic gauge that is defined by the
condition:
\begin{equation}
\partial_{A}\left(  h^{AB}-\frac{1}{2}\eta^{AB}h_{C}^{C}\right)  =0.
\label{gauge}%
\end{equation}

The gravitational constant $G_{D}$ of the higher-dimension theory, that
appears in equation (\ref{linear eq}), should satisfy the relation
$G_{D}=G\left(  2\pi R\right)  ^{\delta}$ to recover the conventional results
of General Relativity at large distances \cite{ADD1,colliders}. Another
requisite is the asymptotic stabilization of the volume of the supplementary
space \cite{antoniadis}.

On the other hand, in short distance, the dominant term of the solution is
independent of the topology of the supplementary space. In the static regime,
the solution of equation (\ref{linear eq}), in this approximation order, is
given by:
\begin{equation}
h_{AB}\left(  \vec{X}\right)  =\frac{16\pi\Gamma(\frac{\delta+3}{2})G_{D}%
}{(\delta+1)2\pi^{\left(  \delta+3\right)  /2}c^{4}}\left(  \int\frac{\bar
{T}_{AB}\left(  \vec{X}^{\prime}\right)  }{\left\vert \vec{X}-\vec{X}^{\prime
}\right\vert ^{1+\delta}}d^{3+\delta}X^{\prime}\right)  , \label{solution}%
\end{equation}
where $\vec{X}$ and $\vec{X}^{\prime}$ are spatial coordinates of the ambient
space. By ignoring the topology, we are taking a lower estimate of the
potential strength. To illustrate this, consider a torus topology, as an
example. As a consequence of the periodicity along each transversal direction
of the brane that is implied by this topology, the resulting potential is
mathematically equivalent to a superposition of potentials produced by a net
of images of the source regularly spread in unfolded extra dimensions
\cite{Kehagias}. Therefore, by considering only the term (\ref{solution}), we
are not taking into account the contribution of all images.

In the context of an interferometry experiment, equation (\ref{solution})
gives the higher-dimensional gravitational potential produced by the phase
shifter, which can be a material plate or an electric field, as we are going
to consider later. In its turn, the coupling of the incident neutron with that
gravitational field can be extracted, in the ray optic approximation, from the
Lagrangian of a test particle with mass $m$ that is moving in the brane:%
\begin{equation}
L=mc\left(  g_{\mu\nu}\dot{x}^{\mu}\dot{x}^{\nu}\right)  ^{1/2}, \label{L}%
\end{equation}
where $\dot{x}^{\mu}$ means the derivative of the particle's coordinates with
respect to its proper time. As the motion is restricted to the brane,
$g_{\mu\nu}$ is the induced metric in $z=0$. For a non-relativistic test
particle, as the slow neutron in the experiment, it follows from (\ref{L})
that the gravitational interaction is described by the potential energy
$U_{G}=m\varphi$, where $\varphi=-h_{00}c^{2}/2$ is the modified gravitational
potential calculated from (\ref{solution}).

\subsection{Baryonic source}

Crossing a material medium, the neutron interacts with the atomic nucleus via
the anomalous gravitational force according to the large extra dimension
scenario. Each nucleus can be treated as a non-relativistic source with an
energy-momentum tensor approximately given by $T_{\mu\nu}=\rho_{N}u_{\mu
}u_{\nu}$, where $\rho_{N}$ is the proper baryonic mass of the nucleus and
$u^{\mu}$ is its four-velocity. In the rest frame of the medium, $u^{\mu
}=c\delta_{0}^{\mu}$ in the first approximation. Thus, it follows that the
gravitational potential of a single nucleus evaluated in a point $\vec{x}$ in
the brane is given by:%
\begin{equation}
\varphi\left(  \vec{x}\right)  =-\hat{G}_{D}\left(  \int\frac{\rho_{N}\left(
\vec{x}^{\prime}\right)  f(z)}{\left(  \left\vert \vec{x}-\vec{x}^{\prime
}\right\vert ^{2}+z^{2}\right)  ^{\frac{1+\delta}{2}}}d^{3}x^{\prime}%
d^{\delta}z\right)  , \label{phi}%
\end{equation}
where, for convenience, we have defined $\hat{G}_{D}=$ $4G_{D}\Omega_{\delta}$
and $\Omega_{\delta}=\Gamma(\frac{\delta+3}{2})/(\delta+2)\pi^{\left(
\delta+1\right)  /2}$. As we have already mentioned, if $f\left(  z\right)  $
is a Dirac delta distribution, the potential is not computable in any interior
point in the case of a codimension greater than one. However, an estimate of
the internal potential can be determined by considering that the brane has a
thickness and that the baryonic mass of the nucleus is distributed along the
extra dimensions according to some regular function $f\left(  z\right)  $,
such as a Gaussian function centered at $z=0$. In the leading order, it is
possible to show that the internal potential is proportional to the baryonic
mass density distribution $\rho_{N}\left(  \vec{x}\right)  $ of the nucleus
\cite{dahia}. Therefore, this internal interaction cannot be distinguished
from the strong interaction between the neutron and the nucleus, which is
described by a semi-empirical potential of Wood-Saxel type \cite{frank,leeb}.

An instructive, although non-rigorous, way to estimate the magnitude order of
the potential (\ref{phi}) is to consider that, due to the mass distribution in
the extra dimensions, the neutron and the source are separated in the
$z-$direction by an effective distance $\sigma$, whose exact value depends on
$f(z)$. At this distance, only a fraction of the source inside a 3-ball of
radius $\sigma$ in the brane parallel directions contributes significantly to
the potential. Thus, the potential will be proportional to $G_{D}\left(
\sigma^{3}\rho_{N}\right)  /\sigma^{\delta+1}$, therefore, proportional to the
density distribution in this approximation as mentioned above.

Perfect silicon crystal interferometers use a slow neutron with a wave-length
$\left(  \lambda_{n}\right)  $ of Angstrom order. Thus, the nuclear
interaction, whose effective action is restricted to the nucleus size, can be
approximated by the Fermi pseudopotential. The average of this potential
energy in the medium is given by \cite{rauch}:%
\begin{equation}
U_{F}=\frac{2\pi\hbar^{2}N}{m}b, \label{U}%
\end{equation}
where the parameter $b$ is the forward scattering length and $N$ is the atomic
density of the material. The effects of the so-called neutron optical
potential (\ref{U}) on the phase shift of the neutron beam can be determined
empirically. According to the available data, the extracted value of $b$ is
roughly proportional to $A^{1/3}$, where $A$ is the atomic mass of the nucleus
\cite{neutrons}. If there is a hypothetical interaction with a gravitational
field like (\ref{phi}), then the measured value of $b$ should contain a small
contribution from its internal part, $b_{G}^{int}$. However, it would be
indiscernible from the nuclear scattering length because of the reasons we
have pointed out previously. The incident neutron may also have other
interactions with the atom which can influence the parameter $b$, but they are
weaker than the nuclear force and can be ignored here, considering our purposes.

The external part of the potential (\ref{phi}) give also a contribution to the
scattering length which, in principle, could be differentiated from the
nuclear scattering length. The exterior gravitational potential in brane
models depends on internal characteristics of the nucleus (its radius, for
instance) even when we assume a spherically symmetric distribution in the
ordinary space. To illustrate this, let us consider the case of five extra
dimensions ($\delta=5$). Modeling the nucleus by a 3-sphere of a certain
radius $R_{N}$, in the ordinary three-dimensional space, with a uniform mass
density, it follows from (\ref{phi}) that the potential of the single nucleus
with total mass $M$ is given by:%
\begin{equation}
\varphi_{ext}\left(  \vec{x}\right)  =-\frac{\hat{G}_{D}M}{\left(  r^{2}%
-R_{N}^{2}\right)  ^{3}}. \label{phi5}%
\end{equation}
This expression is obtained in the thin brane limit. Around the nucleus, the
external potential (\ref{phi5}) is quite different from the potential produced
by a point-like mass and even diverges at $R_{N}$. Considering this, one way
we could estimate $b_{G}^{ext}$, in the Born approximation \cite{leeb}, is
taking
\begin{equation}
b_{G}^{ext}=\frac{m}{2\pi\hbar^{2}}\int_{R_{N}+\sigma}^{\infty}m\varphi
_{ext}\left(  \vec{x}\right)  d^{3}\vec{x},
\end{equation}
where $\sigma$ is the length scale in which the thin-brane limit fails, which
is of the order of the brane thickness as we have already mentioned before. In
the leading order, we find:%
\begin{equation}
b_{G}^{ext}=-\frac{m}{2\pi\hbar^{2}}\left(  \hat{G}_{D}\frac{\pi mM}%
{4R_{N}\sigma^{2}}\right)
\end{equation}
In general, for codimensions $\delta>3$, we have similar results which can be
summarized in a single formula. Writing $G_{D}$ in terms of the Newtonian
constant $G$ and $R$ (the compactification scale of the hidden dimensions), we
find:%
\begin{equation}
b_{G}^{ext}=-\frac{m}{2\pi\hbar^{2}}\left(  \vartheta\frac{R^{\delta}}%
{R_{N}\sigma^{\delta-3}}GMm\right)  \label{bG_ext}%
\end{equation}
where $\vartheta$ is a coefficient whose value depends on the number of hidden
dimensions. The above expression could be rewritten in a way that resembled
what we would get from Yukawa parametrization, provided we reinterpret the
amplification factor as $\alpha=\vartheta R^{\delta-2}/(4\pi R_{N}%
\sigma^{\delta-3})$. This new parameter can reach much higher values in
comparison to the standard $\alpha$, which is just proportional to $\delta$ as
predicted by the original ADD theory for large distances
\cite{Kehagias,adelbergREV}.

In principle, $b_{G}^{ext}$ could be distinguished from nuclear scattering
length due to its peculiar dependence on the atomic mass. Indeed, following
the method described in Ref. \cite{neutrons}, we could check whether the data,
collected from different types of materials, are compatible with an extra
scattering length that is proportional to $M/R_{N}$ and, therefore, to
$A^{2/3}$. However, our ability to establish a clear constraint on the
parameters of the extra dimensions theory is limited by the fact that
$b_{G}^{ext}$ depends also on the nuclear model.

\subsection{Non-baryonic source}

It is possible to obtain empirical bounds that do not depend on nuclear
models, considering a non-baryonic phase shifter as a source for the
gravitational field inside the interferometer. In the experiment designed to
test the Aharonov-Casher effect \cite{cimmino}, neutron beams pass through the
interior of capacitors, accumulating phase shifts due to the interaction
between the electric field ($\vec{E})$ and the neutron's magnetic moment
$\left(  \vec{\mu}\right)  $ imposed by the spin-orbit coupling $\left(
\mu/mc\right)  \vec{\sigma}\cdot\left(  \vec{E}\times\vec{p}\right)  $, where
$\vec{\sigma}$ stands for the Pauli matrices.

In contact with the electric field, the incident neutron does not interact via
nuclear force, but it interacts gravitationally. Indeed, according to General
Relativity, the energy and the stress of the electric field inside the
capacitor produce a gravitational field which affects the motion of the
neutron. The standard theory predicts a negligible effect that is not
detectable within the current precision of the instruments. However, in the
context of modified gravitational theory, such as the large extra dimension
models, the expected amplification of gravity in short distances could be
tested without being masked by the nuclear interaction, even in the case when
the length scale of the anomalous interaction is smaller than the nuclear size.

In the mentioned experiment, the field is approximately uniform in the region
between the electrodes, and its direction, let us say $x_{2},$ is
perpendicular to the direction of the neutron beam ($x_{1}$-axis). In SI
units, the stress-energy tensor of the electromagnetic field is given by:%

\begin{equation}
T_{\mu\nu}^{\left(  EM\right)  }=\epsilon_{0}c^{2}\left(  F_{\mu\lambda}%
F_{\nu}^{\;\lambda}-\frac{1}{4}\eta_{\mu\nu}F_{\alpha\beta}F^{\alpha\beta
}\right)  ,
\end{equation}
where $F_{\mu\nu}$ is the electromagnetic tensor and $\epsilon_{0}$ is the
electric permittivity of the free space. Inside the capacitor, the non-null
components are $F_{20}=-F_{02}=E/c$, where $E$ is the field strength.
Therefore:%
\begin{equation}
T_{\mu\nu}^{\left(  EM\right)  }=\frac{1}{2}\epsilon_{0}E^{2}\left(
\begin{array}
[c]{cccc}%
+1 & 0 & 0 & 0\\
0 & +1 & 0 & 0\\
0 & 0 & -1 & 0\\
0 & 0 & 0 & +1
\end{array}
\right)  .
\end{equation}
This tensor describes an anisotropic stress distribution with an energy
density $u=\frac{1}{2}\epsilon_{0}E^{2}$. In the orthogonal directions of the
field $\vec{E}$, the effective pressures measure $P_{1}=P_{3}=P_{\perp}=u$.
But, in the parallel direction, the field configuration gives rise to a
tension $P_{2}=P_{\parallel}=-u$. The average pressure satisfies the usual
state equation of radiation, $\hat{P}=u/3$.

Taking into account that $T^{\left(  EM\right)  }$ has a null trace, it
follows from (\ref{solution}) that the gravitational interaction with the
incident neutron can be described by the potential:
\begin{equation}
\chi\left(  \vec{x}\right)  =-\frac{\left(  \delta+2\right)  }{\left(
\delta+1\right)  }\frac{\hat{G}_{D}}{c^{2}}\left(  \int\frac{u\left(  \vec
{x}^{\prime}\right)  f(z)}{\left(  \left\vert \vec{x}-\vec{x}^{\prime
}\right\vert ^{2}+z^{2}\right)  ^{\frac{1+\delta}{2}}}d^{3}x^{\prime}%
d^{\delta}z\right)  , \label{Qui}%
\end{equation}
which is greater than the potential produced by a non-relativistic source with
the same energy density by a factor $\left(  \delta+2\right)  /\left(
\delta+1\right)  $.

As we have already mentioned, in the zero-width brane idealization, the
potential diverges in any point $\vec{x}$ where $u$ is non-null, in the case
of $\delta\geq2.$ However, this internal potential can be calculated in a
thick brane scenario, admitting that the confinement of the electric field in
the brane is described by any regular and normalized distribution $f\left(
z\right)  $.

In order to simplify the calculation of the internal potential, let us assume
the realistic hypothesis that $R$ is much smaller than the distance between
the electrodes ($d=1.54$ $%
%TCIMACRO{\unit{mm}}%
%BeginExpansion
\operatorname{mm}%
%EndExpansion
$). Then, in any interior point $\vec{x}$, away from the capacitor's boundary,
the major contribution for the potential comes from a portion of the source
contained in a region $B_{3}(R)$, which corresponds to a spherical
neighborhood of radius $R$ in the ordinary three-dimensional space center at
$\vec{x}$. For $\delta>2$, in the leading order, we get from (\ref{Qui}):
\begin{equation}
\chi\left(  \vec{x}\right)  =-\frac{4\pi\zeta\hat{G}_{D}}{\varepsilon
^{\delta-2}}\frac{1}{2}\frac{\epsilon_{0}E^{2}}{c^{2}}, \label{Qui principal}%
\end{equation}
where the coefficient $\zeta$, in terms of the gamma-function ($\Gamma$), is:%

\begin{equation}
\zeta=\frac{\sqrt{\pi}}{8}\frac{\delta\left(  \delta+2\right)  \Gamma
(\frac{\delta-2}{2})}{\left(  \delta+1\right)  \Gamma(\frac{\delta+1}{2})}.
\end{equation}
The next correction term has a relative order of $\varepsilon/R$, at least.
The potential $\chi$ also depends on the parameter $\varepsilon$ that is
defined as:
\begin{equation}
\frac{1}{\varepsilon^{\delta-2}}=\frac{2}{\delta}\int\frac{f(z)}{z^{\delta-2}%
}d^{\delta}z,
\end{equation}
which is the average of the function $1/z^{\delta-2}$ with respect to the
distribution $f(z)$. As pointed out in Ref. \cite{dahia3}, in the leading
order, the gravitational field produced by localized sources inside the brane
does not depend on many details of the confinement mechanism, but on a
specific statistical moment (the expected value of $z^{2-\delta}$) of the
field distribution in the supplementary space.

The part of the source which lies outside the region $B_{3}(R)$ gives a
contribution for the resulting potential that has a relative order of $\left(
\varepsilon/R\right)  ^{\delta-2}$.

In the next section, we are going to discuss the effect of this potential on
the incident neutron.

\section{Constraints from the neutron interferometry}

For the sake of simplicity, let us consider a minor change of the AC
experiment. Let us admit that only one of the partial beams passes through the
region filled with the electric field, while the other beam is shielded.
Inside the capacitor, the neutron will interact with the electric field
through the gravitational anomalous interaction too. This extra interaction
will provide an additional phase factor to the beam's wave function.
Considering that it is a short-range interaction, the second beam will not be
affected by the anomalous potential. Thus, the relative phase between the two
paths will be given by \cite{rauch}:%
\begin{equation}
\Delta\Phi=\frac{1}{\hbar}\int_{C}\Delta\vec{p}\cdot d\vec{s}, \label{phase}%
\end{equation}
where $\hbar$ is the reduced Planck constant and $\Delta\vec{p}$ is the
variation of the neutron linear momentum caused by the anomalous gravitational
interaction in relation to the linear momentum of the free neutron. The
integration is performed along the path of the first beam inside the
capacitor. In this approximation, we are neglecting the contribution from the
potential outside the medium.

The variation of the linear momentum can be determined from the energy
conservation and can be expressed in terms of the wavelength of the incident
neutron. Along the direction of motion $(x_{1}$-axis, inside the capacitor),
we find $\Delta p_{1}=m^{2}\chi\lambda_{n}/h,$ where $\chi$ is constant and is
given by (\ref{Qui principal}) in the leading order. Therefore, the phase
difference acquired by the beam after traversing the capacitor of length $L$
is:%
\begin{equation}
\Delta\Phi=\frac{m^{2}}{h^{2}}\left(  \frac{4\pi^{2}\zeta\hat{G}_{D}%
}{\varepsilon^{\delta-2}}\epsilon_{0}\frac{E^{2}}{c^{2}}\right)  \lambda_{n}L.
\label{PhaseG}%
\end{equation}
As we have pointed out before, interestingly, the result (\ref{PhaseG}) can be
formally derived from a Yukawa parametrization too. If we take $\lambda=R$ and
reinterpret the dimensionless Yukawa parameter as $\alpha=\left(  2\pi\right)
^{\delta}\zeta\Omega R^{\delta-2}/\varepsilon^{\delta-2}$, then the formula
above could be rewritten as $\Delta\Phi=G\alpha\lambda^{2}m^{2}\epsilon
_{0}E^{2}\lambda_{n}L/h^{2}c^{2}$, i.e., in the same form that would be
obtained from an anomalous gravitational field described by the Yukawa
parametrization. Thus, considering that the Yukawa parametrization can
describe modifications of gravity with other physical origins besides hidden
dimensions, then in order to be more generic as possible from the
phenomenological point of view, we are going to express our results in terms
of the Yukawa parameters hereafter.

In the AC experiment \cite{cimmino}, the field strength is $E=30%
%TCIMACRO{\unit{kV}}%
%BeginExpansion
\operatorname{kV}%
%EndExpansion
/%
%TCIMACRO{\unit{mm}}%
%BeginExpansion
\operatorname{mm}%
%EndExpansion
$, $L=2,53%
%TCIMACRO{\unit{cm}}%
%BeginExpansion
\operatorname{cm}%
%EndExpansion
$ and the neutron wavelength is $\lambda_{n}=1.477%
%TCIMACRO{\unit{\U{212b}}}%
%BeginExpansion
\operatorname{\text{\AA}}%
%EndExpansion
$. The predicted phase shift $\Phi_{AC}=1.50$ mrad is compatible with the
measurements within an error of the order of $\eta\sim$ 10$^{-3}$rad
\cite{cimmino}. Therefore, additional effects from any hypothetical
interaction could not be greater than $\eta$.

In principle, the gravitational phase shift, which depends on $E^{2}$, can be
distinguished from the AC shift, which is proportional to $E$. Thus, it seems
reasonable to expect that we can get constraints for the anomalous
gravitational interaction from the analysis of an experiment of this kind. In
order to make an estimate, let us assume that the empirical procedure of
testing the modified gravity produced by the capacitor electric field has a
precision of the same order of $\eta$. Thus, it follows from (\ref{PhaseG})
that the corresponding Yukawa parameter should satisfy the upper limit:%

\begin{equation}
\alpha\lambda^{2}<0,26\times10^{20}%
%TCIMACRO{\unit{m}}%
%BeginExpansion
\operatorname{m}%
%EndExpansion
^{2}\left(  \frac{\eta}{10^{-3}%
%TCIMACRO{\unit{rad}}%
%BeginExpansion
\operatorname{rad}%
%EndExpansion
}\right)  \left(  \frac{%
%TCIMACRO{\unit{\U{212b}}}%
%BeginExpansion
\operatorname{\text{\AA}}%
%EndExpansion
}{\lambda_{n}}\right)  \left(  \frac{%
%TCIMACRO{\unit{cm}}%
%BeginExpansion
\operatorname{cm}%
%EndExpansion
}{L}\right)  \left(  \frac{30%
%TCIMACRO{\unit{kV}}%
%BeginExpansion
\operatorname{kV}%
%EndExpansion
/%
%TCIMACRO{\unit{mm}}%
%BeginExpansion
\operatorname{mm}%
%EndExpansion
}{E}\right)  ^{2}, \label{NI bound}%
\end{equation}
which is valid for $\lambda<10^{-4}%
%TCIMACRO{\unit{m}}%
%BeginExpansion
\operatorname{m}%
%EndExpansion
$. Although this bound is very weak compared to the traditional constraints
based on baryonic sources \cite{neutrons}, we should emphasize that
(\ref{NI bound}) can be extended to length scales smaller than nuclear size.
Moreover, it is also valid for large extra-dimensions models, provided the
Yukawa parameter be appropriately interpreted as discussed before.\qquad
\qquad\qquad\qquad

\section{Post-Newtonian potentials}

Strictly speaking, the potential $\chi$ cannot be interpreted as a
modification of the Newtonian potential since its source is not the matter's
rest mass. Instead, it should be considered as a short-distance deviation of a
post-Newtonian correction potential, given that the electric energy acting as
a source of a gravitational field has no correspondence in the Newtonian theory.

In the weak-field regime, alternative metric theories can be distinguished by
means of parameters that work as effective gravitational couplings related to
post-Newtonian potentials. In the standard PPN formalism, there are ten
parameters \cite{ppn}. Two of them, $\beta_{3}$ and $\beta_{4}$ (following the
notation of Ref. \cite{MTW}), are of special importance here. They measure how
much gravity is produced by internal non-baryonic energy and by pressure in
comparison to the General Relativity predictions, respectively.

The potential $\chi$ is influenced by a combination of the energy density
$(u)$ and the average pressure $(\hat{P})$ associated with the electric field.
Thus, in accordance with this more general formalism, in which a broader class
of metric theories could be considered (not only extra dimensions models), the
parameter $\alpha$ should be replaced, in the constraint equation
(\ref{NI bound}), by a combination of two Yukawa parameters associated with
short-distance modifications of the post-Newtonian parameters $\beta_{3}$ and
$\beta_{4}$.

In general, the post-Newtonian parameters are independent, and their values
are to be determined from phenomenology. However, for a restricted class of
metric theories that satisfy the full global conservation laws and that are
free of preferred spatial position, $\beta_{3}=1,$ automatically
\cite{ppn,MTW}. This implies that all kinds of energy have the same capacity
of curving spacetime. On the other hand, the gravitational coupling associated
with the pressure is not fixed in this class of theories, but should satisfy
the relation \cite{ppn,MTW}: $\beta_{4}=\gamma$, where $\gamma$ is another
post-Newtonian parameter, that is associated with the curvature of the pure
spatial sections of the spacetime.

The prediction of GR theory, $\gamma=1$, has been confirmed by tests such as
the Cassini experiment that investigated the time-delay and the deflection of
radio waves under the influence of the gravitational field of the Sun
\cite{gamma}. The empirical value, $\gamma=1+(2.1\pm2.3)\times10^{-5}$, is the
most stringent bound for that parameter in the length scale of the solar
radius \cite{gamma}.

In the microscopic domain, this parameter has been investigated too, by
examining possible effects of short-range modifications of gravity through the
analysis of the gravitational spin-orbit coupling, which depends on $\gamma$
\cite{fish}. The MTV-G experiment, for example, is based on the spin
precession of electrons that are scattered by heavy nuclei \cite{mtvg,mtvg1}.
The absence of any anomalous signal in the empirical data establishes some
bounds on short-distance deviations of the parameter $\gamma$
\cite{mtvg,dahia4}.

There are also constraints extracted from the spectroscopy of the hydrogen
atom. The fine structure of $P$-states is influenced by the spin-orbit
coupling mediated by the gravitational field produced by the nucleus. The
analysis of $2P_{1/2}-2P_{3/2}$ transition put independent constraints on the
short-range behavior of that post-Newtonian parameter too \cite{dahia4}.

It is interesting to compare (see Figure 1) the MTV-G and spectroscopic bounds
with the constraint extracted from neutron interferometry, since, for a
restricted class of metric theories, the coefficient $\alpha$ in
(\ref{NI bound}) can be viewed effectively as the Yukawa parameter related to
the short-distance modifications of $\gamma$, i.e., $\gamma\left(  r\right)
=\left(  1+\alpha e^{-r/\lambda}\right)  $.%

%TCIMACRO{\FRAME{dtbpFU}{2.1447in}{2.1447in}{0pt}{\Qcb{\QTR{small}{Figure 1.
%Neutron Interferometric (NI) bounds (obtained in this paper) on the Yukawa
%parameter }$\alpha$ \QTR{small}{ related to the short-range modifications of
%the Post-Newtonian }$\gamma-$\QTR{small}{parameter in comparison to the MTV-G
%and spectroscopy limits (extracted from \cite{mtvg,mtvg1,dahia4}). The region
%above the lines are excluded.}}}{\Qlb{Figure}}{figure_neutron.bmp}%
%width 2.1447in;  height 2.1447in;  depth 0pt;  original-width 2.6671in;
%original-height 2.6671in;  cropleft "0";  croptop "1";  cropright "1";
%cropbottom "0";  filename 'Figure_Neutron.bmp';file-properties "XNPEU";}}}%
%BeginExpansion
\begin{center}
\includegraphics[height=2.1447in,
width=2.1447in
]%
{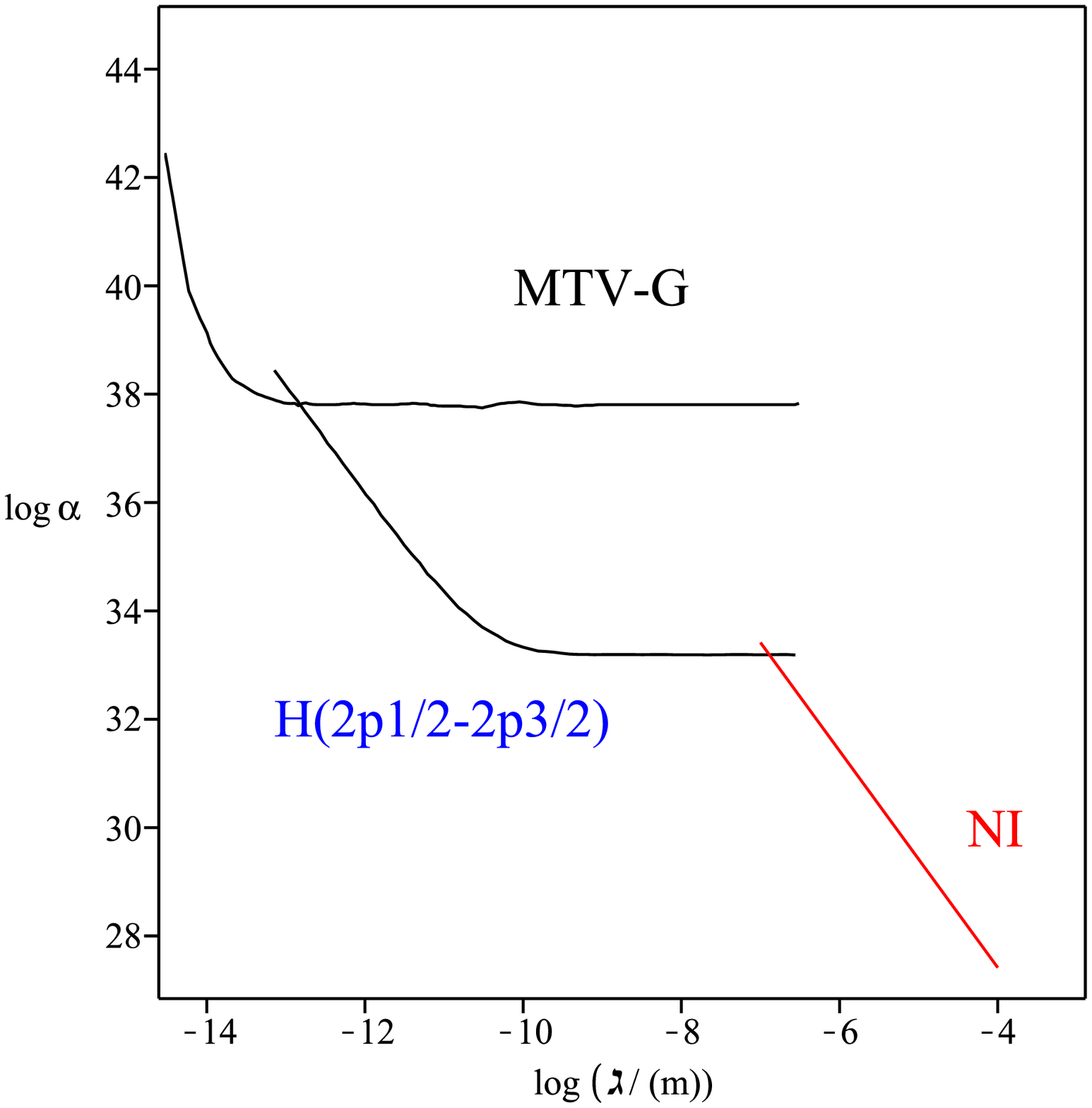}%
\\
{\protect\small Figure 1. Neutron Interferometric (NI) bounds (obtained in
this paper) on the Yukawa parameter }$\alpha$ {\protect\small  related to the
short-range modifications of the Post-Newtonian }$\gamma-$%
{\protect\small parameter in comparison to the MTV-G and spectroscopy limits
(extracted from \cite{mtvg,mtvg1,dahia4}). The region above the lines are
excluded.}%
\label{Figure}%
\end{center}
%EndExpansion

The neutron interferometry yields the strongest upper limits for
amplifications of the post-Newtonian $\gamma-$parameter in the range
$1.4\times10^{-7}%
%TCIMACRO{\unit{m}}%
%BeginExpansion
\operatorname{m}%
%EndExpansion
$ and $10^{-4}%
%TCIMACRO{\unit{m}}%
%BeginExpansion
\operatorname{m}%
%EndExpansion
$.

\section{Final remarks}

Large extra dimensions theories are often cited as motivation in the search
for modifications of gravity on short-distance scales in several laboratory
tests. However, the thin brane models cannot be probed by experiments based on
neutron interferometry straightforwardly, since the internal potential of a
material sample, playing the role of a phase shifter, is not computable in
scenarios where the brane has no thickness and the number of codimensions is
greater than one.

Therefore, when interferometric constraints are expressed in terms of Yukawa
parameter $\alpha$, it is not clear what physical quantity related to the
higher-dimensional theory could, in fact, be bound by the empirical data.

The ADD model predicts that, far from a pointlike source, the Newtonian
potential is corrected by an additional Yukawa term. Considering a
supplementary space with a flat torus background and a model with massless
radion, the Yukawa parameter is $\alpha=2\delta$ \cite{Kehagias}. This
interpretation is valid, for example, in torsion-balance tests, since the
bodies used in the experiment are spatially separated. Thus, taking into
account the finite size effects of the interacting bodies, constraints on the
compactification radius of large extra-dimensional models can be extracted
from the torsion-balance experiments employing the Yukawa parametrization
\cite{Hoyle}.

However, in the interferometry experiment, the neutron is in contact with the
material (the phase shifter) in a certain interval of its path. Therefore, the
Yukawa parametrization is not appropriate to describe the effects of hidden
dimensions on the neutron's phase factor. By the way, if we inadvertently
employ the above interpretation for $\alpha$ with the purpose of analyzing the
interferometric bounds, the upper limits we obtain for the number of hidden
dimensions are practically irrelevant.

Inside the material, the power-law parametrization would be the adequate one
to study modifications of gravity in the context of the braneworld scenario.
However, it leads to divergence problems in the calculation of the internal
gravitational potential.

This difficulty can be circumvented in the context of a thick brane scenario.
Indeed, considering that the localized fields have a regular distribution
inside the thick brane, we have calculated the forward scattering length,
$b_{G}$, associated with the higher-dimensional gravitational interaction
between the neutron and the nucleus. As we have seen, the part of $b_{G}$ that
can be distinguished from the nuclear scattering length $(b_{G}^{ext})$ is
proportional to ratio $R^{\delta}/\left(  R_{N}\sigma^{\delta-3}\right)  $,
which involves the compactification radius $R$, the nucleus radius $R_{N}$ and
$\sigma$, a parameter of the order the brane thickness that is associated with
a statistical moment of the distribution that describes the localized field
inside the brane.

The explicit determination of $b_{G}^{ext}$ allows us to recognize the
higher-dimensional quantity that is subjected to the empirical constraints put
by neutron interferometry. However, the parameter $b_{G}^{ext}$ depends on the
nuclear model. Hence, in order to obtain constraints free of this dependence,
we were led to consider experiments where the phase shifter is non-baryonic,
as in the experiment that measures the Aharanov-Casher shift.

In the context of PPN formalism, the AC experiment can be seen as a test of
the behavior of Post-Newtonian $\gamma-$parameter on the short-length scale
for a class of metric theories. In comparison to other empirical constraints,
extracted from the MTV-G experiment and from the $2P_{3/2}-2P_{1/2}$
transition in the hydrogen atom, we find that the limits imposed by neutron
interferometry on deviations of that parameter are the most stringent in the
length scale between $1.4\times10^{-7}%
%TCIMACRO{\unit{m}}%
%BeginExpansion
\operatorname{m}%
%EndExpansion
$ and $10^{-4}%
%TCIMACRO{\unit{m}}%
%BeginExpansion
\operatorname{m}%
%EndExpansion
$.

\begin{acknowledgement}
J. M. Rocha thanks CNPq for financial support.
\end{acknowledgement}

\end{document}